\documentclass[10pt,a4paper]{article}
\usepackage[utf8]{inputenc}
\usepackage[T1]{fontenc}
\usepackage{amsmath}
\usepackage{amsfonts}
\usepackage{amssymb}
\usepackage{graphicx}
\usepackage{fullpage}

\begin{document}
\title{Quantum state collapse on a Riemann-Hilbert space modulo observation}	
\author{Jose A. Pereira Frugone \\ japfrugone@yahoo.com}

\maketitle

\textbf{Abstract} \\
In a previous work we constructed a new kind of moduli background space by identifying regions of space-time where an observation of space-time is implied. We called it Observation Modular space (OM-space). Quantum Mechanics (QM) on this moduli space gets mapped into a very rich and highly non trivial dual Number Theory which we call Observation Modular Quantum Mechanics (OM-QM). In this work we extend the scope of this modularization to include observations of quantum states. We call the resulting extended space Observation Modular Riemann-Hilbert space (OM-RH space). It has the mathematical structure of a Riemann Surface. We find the OM-QM analogue of quantum base states and mixed states. This allows us to find the OM-QM analogues to the quantum State Reduction postulate and the Born rule. The OM-QM analog to quantum state collapse turns out to be totally deterministic and unitary in OM-RH space. It is shown to be equivalent to an Elliptic Curve Encryption-decryption protocol. Finally we obtain the OM-QM analog of entangled quantum states. As an example we apply this to the OM-QM interpretation of the EPR experiment.

\section{Introduction}
In a previous work \cite{frugone2023quantum} we answered the following question: What remains of Quantum Mechanics when we identify (modularize) regions of the space-time background where observation or measurement is implied ? We found that residual Theory (which we called Observational Modular Quantum Mechanics or OM-QM) using as guiding principle the procedure described in these three steps :

\begin{itemize}
	
	\item[\bf{step 1}] Take a valid Quantum Mechanical statement (an equation, a definition, a postulate). The OM-QM corresponding statement can be constructed via the following operations:\\
	\item[\bf{step 2}] Replace any instance of the Planck constant h by the quantity $\zeta_0 4 \pi^2$ (where $\zeta_0$ is the Planck Length). Or equivalently, replace $P_0$ (the Planck Momentum) in the QM side by the number $4 \pi^2$ \\
	\item[\bf{step 3}] Identify all regions in the background space-time where a quantum phase comparison (that is measurement or observation) is implicit in the quantum statement (up to homotopy classes). Keep the form of the QM equations covariant in the transformation. Then find the effect of such transformation in all elements in the quantum statement.
	
\end{itemize}

We call this procedure the OM-correspondence and the moduli space in Step.3 the Observation Modular space (OM-space). We proved that OM-QM is a very interesting and highly non trivial Number Theory. This Physical modularization transforms some problems which are hard or obscure in the Qantum Mechanical (QM) side to become more tractable or simpler to model in the OM-QM side. Off course OM-QM is a purely Mathematical framework but due to the way it is constructed it conserves a lot of information about the QM source side. It also adds its own Mathematical insights which in turn sheds some light on those mysterious or hard aspect of the QM side. In Mathematics we often see a similar simplification or additional insight when one applies modular techniques to some hard Mathematical problems. We suggest OM-QM is a Physics example of the same kind of simplification and insight coming from a Physical modularization in this case.  \\ \\ 

In the first section of this paper we present a recapitulation of the main results in \cite{frugone2023quantum}. This first section can be skipped if the reader is familiar with content of that work. For this paper we will start by emphasizing that in \cite{frugone2023quantum} we focused exclusively in Observation in relation to the background space-time on which we build QM. The modularization introduced in that work was exclusively identification of regions of space-time where Observation was implicit or explicit. In this present work we will focus on another kind of Observation, the Observation of quantum states. We will add a new step (Step 4.) to the three ones mentioned above for the OM-correspondence. This new step  says that we should identify any quantum base states to the one which has been observed (according to an equivalence relation we will define). This sets up what we will call an Observation Modular Hilbert space (OM-H space). \\ \\

 If we are talking about Observation of mixed base quantum states then we have to consider (from the start) the so called State Collapse principle of Quantum Mechanics. This fundamental principle states that after Observation or Measurement on a system represented by a mixed state this quantum state collapses (or gets reduced) to one single base state which is selected in a totally random way. This is a non-Unitary process which is one of the most mysterious and hard problems in QM. We will see that in OM-QM we can define the equivalent to base states, mixed states and that OM-QM maps the quantum state collapse process to a much more tractable and insightful Mathematical problem.  We find an explicit OM-QM Mathematical mechanism representing this process in the OM-QM side and a visualization aid for it. This visualization aid is analog to a Key-Cylinder opening mechanism. This process is totally Unitary if we consider it happening in the full OM-RH space. However it is chaotic if viewed only on OM-space. It also give us an OM-QM version of the Born rule for which we will find an explicit form. \\ 
  A good portion of this work will be devoted to exploring the mathematical structure of this OM-RH space. This structure turns out to be the one of a Riemann Surface \cite{teleman2003riemann} with genus equal to the number of geometrical constraints in the QM side. This allows us to make use of the rich machinery of Riemann Surfaces and other Mathematical subjects related to it for the analysis of OM-QM. \\ \\ 
 
 In essence, we are suggesting that the OM-QM Theory constructed only with the first three steps in the procedure is suitable for any Physical process where no Quantum Measurement is involved. Only Observations of space-time and Unitary evolution of quantum states are covered by those three steps. However when we start considering  quantum Observations we will need to invoke that new Step 4. of the full OM-correspondence. Using this as guiding principle we will be able to make an explicit model of the fuzzy center of the OM-Knot structure found in \cite{frugone2023quantum} and get its relation to the base states in the OM-RH space. We will translate Step 4. into concrete Number Theory and Geometry models. A central insight to do this will come from the differentiation of the parts of the OM-Knot related to the space-time modularization and the parts coming from the Hilbert space modularization. We will see that the OM-Knot petals coming from the modularization of space-time are fundamentally different to the loops coming from the center minimal cell fuzziness. For the former, their length is dependent on the metric distance in the source space-time. The number of the later is dependent on the length of the OM-petal loop and in the symmetries in the QM system.  They are however topological in origin and their length can be contracted close to the minimal cell in OM-space. The randomness we observe in the quantum state collapse gets explained and modeled explicitly by how these different loops interplay in the micro structure of the OM-RH space.  We will see that when the measurement process is considered in the full OM-RH space the process is not random at all. The structure of OM-RH space as a Riemann Surface will give us an explicit form for an OM-mixed state as the Weirstrass function \cite{teleman2003riemann} over that Riemann Surface. The known  differential equation for the Weirstrass function defines the dynamics of the OM-state in the process of reduction to a single OM-base state.  \\ \\ 
 We will define the OM-QM equivalent of entanglement between two quantum particles and visualize it via an interconnected double Key-cylinder mechanism. The details of this visualization shows that in an OM-two-particle entangled system one of the OM-key  opens the cylinder of the other particle and vice-versa. This allows us to interpret the mechanism of destruction of the entanglement in an measurement of one of the particles as an  Elliptic Curve Encryption (ECC) decryption protocol \cite{gajbhiye2011survey} \cite{menezes2018handbook}. We will get an explicit formula for the OM-two-particle entangled system as the genus-2 Weirstrass function \cite{england2012generalised}. In general a quantum system with k geometrical constraints and r entanglements will have an OM-QM mixed state given by the Weirstrass function over a Riemann surface of genus k and r bridges.  \\ 
 
 As an application of this framework we will analyze the OM-correspondent to the Einstein-Podolski-Rosen (EPR) experiment. We find an explicit OM-QM mechanism for the observed quantum results.\\ \\
 
 Readers familiar with the results of our previous work (\cite{frugone2023quantum}) can skip the next section and go straight to section 3.

\section{Recapitulation and precision on Observation Modular Quantum Mechanics OM-QM}

In \cite{frugone2023quantum} we constructed a very specific moduli space for which the reduction of QM on it is a highly structured and non-trivial Number Theory. The modularization is done by identifying regions of space-time where Observation represented as quantum phase comparison (Observation) is explicit or implied. We called this moduli space Observation Modular space (OM-space) and the version of QM reduced over this space was called Observation Modular Quantum Mechanics (OM-QM). For this present paper it is crucial to keep in mind that in that previous work we focused only in Observation related to space-time. Although one could define such modularization in apparently infinite ways, we used a very specific parametrization of the target theory. Every statement in Quantum Mechanics is mapped into a corresponding Number Theoretical statement  which results from replacing the Plank Constant (h) with the quantity  $\zeta_0 4 \pi^2$ (where $\zeta_0$ is the Planck Length). \\
Another way to say the same is that we constructed a correspondence between a Quantum Mechanical statements and a certain Number Theory statement which results from replacing the Planck's momentum, defined as 
\begin{equation}\label{Eq1}
P_0 = \frac{h}{\zeta_0} 
\end{equation} 

 with the number $4 \pi^2$ and transforming Space-Time in the aforementioned way. We called this reparametrization the $P_0$-$4 \pi^2$ correspondence. We will call the full process of transforming a QM statement into its OM-QM dual the OM-correspondence. This procedure is defined by the three Steps mentioned in previous section of this paper. 

In the rest of the paper we will notate with a tilde the OM-QM correspondent of any quantum quantity. Then by definition we have 

\begin{equation}\label{Eq2}
\tilde{P_0} = 4 \pi^2 
\end{equation}   

Also, every time we add the prefix OM on a quantity or object "X" we mean the Observation Modular version of this object. That means, OM-X is the OM-QM mathematical object on which the QM side object X gets mapped under the OM-correspondence.
 
Step 2. and 3. assumes we keep in this OM-correspondence the  relationships between source and target elements, a sort of covariance. The OM-correspondence is  not a transformations under which QM is covariant. Examples of the covariance transformation of QM are the Lorentz transformations for the background (for relativistic QM), SU(2) representations for spin parts of the state vector and eventually Gauge transformations if Gauge Fields would be present. Therefore, if we want to impose covariance under our space-time modular transformation the way the QM elements transform must be different than in the way they transforms under its covariance transformations. We must assume that all elements in the QM equation must transform in general in order to keep covariance. Elements as Mass, Gauge Charges (electric or other), Spin and the state vectors will all change in these transformations. One of the main goals in our previous work was finding the OM-QM correspondents of all those basic observables, elements or parameters. \\
The construction of OM-space was done using an operational analogue of Step 3. We found it was enough for an operational definition of OM-space if we could define how the Feynman Amplitude changes in the OM-correspondence. The change of the Feynman paths under OM-correspondence is described in Figure 1

\begin{figure}[h]
	\centering
	\includegraphics[width=0.8\linewidth]{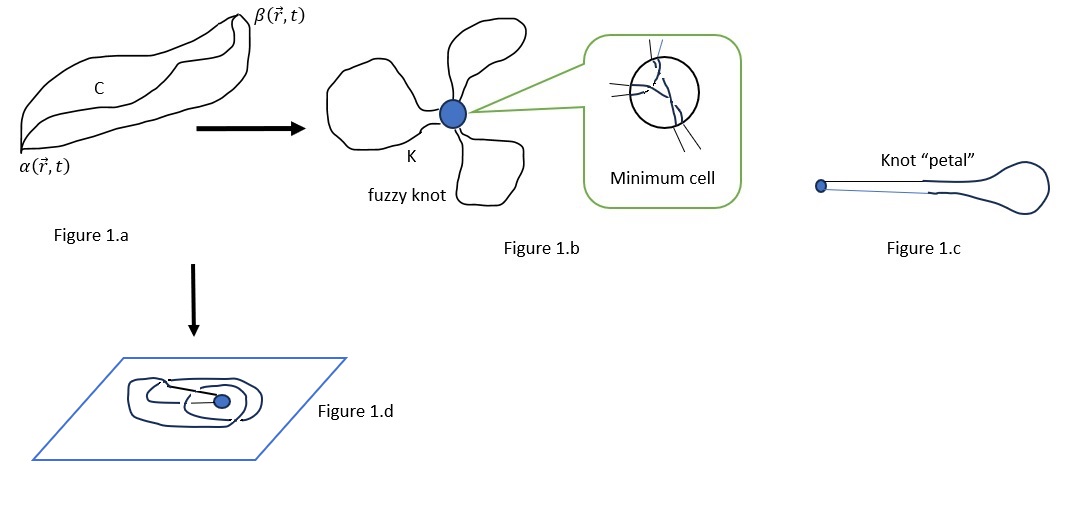}
	\caption[Figure 1]{OM-space elements}
\end{figure}

Here we make a crucial observation about the end points of the set of paths $ \mathbf{C} $ (the start classical space-time point $ \alpha $ and end point $ \beta $ ). There is a quantum phase comparison assumed in each of them. Therefore following Step 3 we can identify both end points  $ \alpha $ and  $ \beta $ up to a cell of minimal size in space-time (a pixel given by Heisenberg's uncertainty relation). Going forward, we will consider that minimal size pixel to be a Planck Length-Time radius ball in 4 dimensions. Crucial to this present work is to recognize that in \cite{frugone2023quantum} we could hide all the quantum details inside this "fuzzy" center and consider only what happened outside of it. We didn't really needed that detail in all that previous work except for describing the global properties of OM-space. In this new work we will focus precisely in that central region. For reasons which will be explained later ignoring the central region in \cite{frugone2023quantum} was consistent with the kind of Observations we were modularizing. The net result of this reduction of number of paths due to topological equivalence and to the identification of the end points as described above is that the set of paths $ \mathbf{C} $ gets mapped to a finite set of closed loops united in a fuzzy way on that minimal size pixel of space-time. \\ \\ 

We notice in Figure 1.c the way the transformed end parts of the paths "approach" the minimal size cell. They enter the region close to the minimal cell in an almost parallel way. We will call that kind of closed loops the OM-petals. In any case the resulting general topology will be always the one of a Knot. Going forward we will call it the OM-Knot. \textbf{Here we want to emphasize something which was not evident yet during our previous work but will be important in this present work. The OM-Knot petals constructed from space-time modularization have a imprint of the metric separation between the events $ \alpha $ and $ \beta $. They are not totally contractible to the minimal central cell via homotophy. This is even more evident when we will have experimental setups in the QM side having diverse geometrical or topological constraints. }\\ \\ 

From this we get the following OM correspondence for one  position coordinate

\begin{equation} \label{Eq3}
\tilde{x}=N \zeta_0
\end{equation}

and the expression of the OM-QM version of the quantum momentum operator 

\begin{equation}\label{Eq4}
\tilde{p}=i 2 \pi \dfrac{d \tilde{\Psi}(N) }{dN}
\end{equation}

where $ \tilde{\Psi}(N) $ is the OM-QM counterpart of the Wave Function.  \\ \\

There is another kind of loops generated in the OM-correspondence. Due to the freedom of winding around the minimal cell of the resulting loop described above. In the modularization transformation, the length of the paths connecting the extreme points can transform into a integer multiple of the minimal cell space size (the Planck  Length)
 
\begin{equation} \label{Eq5}
n \zeta_0
\end{equation}

This variable n is (at this moment) independent and in general different than N. \\ \\

In this paper we will see that this topological way of introducing this central variable n hides a lot of Physical baggage which was not nECCssary to understand for the results in \cite{frugone2023quantum} but is central for the scope of the present work. Here will need to understand a very precise and physical origin of this variable n as representing the information hidden in the fuzzy center of the OM-Knot. Most important, these loops are fundamentally different in nature to the petals we mentioned before. These n loops are topological in nature. They can be contracted very close to the central minimal cell. As we mentioned before, except for the visualization of OM-space and its global properties we completely ignored this variable n in the previous work \cite{frugone2023quantum}.  \\ \\

We also found the OM-QM correspondent for the Speed of Light (c) 

\begin{equation}\label{Eq6}
\tilde{c} = - 2i \pi
\end{equation}

In the following figure we see a visualization of this medium scale structure of OM-space.
\\ \\

\begin{figure}[h]
	\centering
	\includegraphics[width=0.7\linewidth]{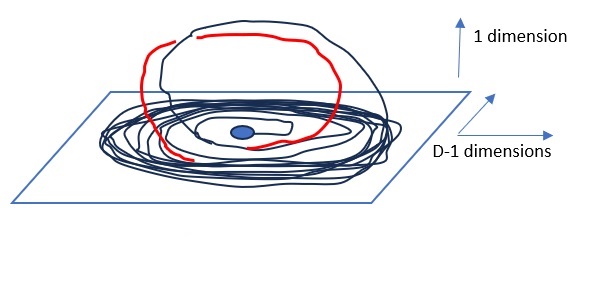}
	\caption[Figure 2]{Rosller-like visualization of medium scale structure of OM-space}
\end{figure}

 It is very similar to an object known in the area of Dynamical Systems and Chaos as a Rosller attractor \cite{rossler1976equation} \cite{letellier2006inequivalent}. To be precise this is just a visualization analogy. The real target object has a very close resemblance of the Rosller attractor but its real structure can have some variations as we will see later. It is high dimensional in nature which will force us to produce several visualization aids for learning its properties. Going forward we are going to use the name Rosller-like geometry to describe this middle structure of the OM-pace. It shows a flow of loops around a central region in an almost two dimensional ring with re-injection flow along a perpendicular direction. The resulting object has a dimension between 2 and 3 with a fractal geometry. Locally this kind of strange attractors behave as a Baker Map. That is, each loop around central region stretches the complement space volumes in one of the coupled dimensions while expanding them in the non coupled directions. The net effect is a volume preserving transformation for each cycle of orbits flow around the central region. A known characteristic of this kind of Baker flows is that the separation between distances ($ d_N $) between points in the flow increases in an universal exponential fashion involving the Lyapunov ($ \lambda $) coefficients of the loops flow 
 
 \begin{equation}\label{Eq7}
 d_{N+1} \propto d_{N} K \exp{(\sqrt{\pi \lambda })}
 \end{equation}
 
 where for a Baker type map (locally a logistic like map), the Lyapunonv coeficient is equal to the Feigenbaund Constant ($ \delta = 4,66920160910299.... $ )
 
 The proportionality constant K depends on the square of the Hausdorff dimension of the Rosller fractal. If the attractor is embedded in a three dimensional space then this proportionality constant will be the square of a number between 2 and 3 and rather close to 3 (the dimension of a locally logistic fractal embedded in 3D space). This will be important in another section later in this paper.
 
 It is worth noting that every observable of the system depending only on the scale we observe the system follows the same scaling universal behavior. For example, in many systems one of those properties depending only on the observation scale is Entropy.  Based on this Rosller-like structure of OM-space were able to find an explicit prediction of the value of the Fine Structure Constant which is related to the aforementioned Lyapunov factor when we fix the dimension of the Rosller fractal to be precisely D=2.974283562752.....

\begin{equation}\label{Eq8}
\tilde{\alpha}^{-1} = D \sqrt{\exp{(\sqrt{\pi \delta})}}
\end{equation}

\textbf{For this present paper it will be important understanding that $\alpha$ being a quantum amplitude it can be interpreted also as the volume of some geometrical object. Volumes in relation to the Rosller object can be calculated as certain multiples of $\alpha$.} \\ \\

We found the OM-QM counterpart of the free particle Dirac Equation
 
   \begin{equation}\label{Eq9}
 \tilde{s} \dfrac{\delta}{\delta N}  \tilde{\Psi } (N) = \tilde{m}(N) \tilde{\Psi } (N)
  \end{equation} 
 
and the OM-QM equivalent of spin

\begin{equation}\label{Eq10}
\tilde{s} = \pm (i-1) 
\end{equation}

 Therefore the general solution to Eq.\ref{Eq9} is 
 
 \begin{equation}\label{Eq11}
 \tilde{\Psi } (N) =  e^{\frac{\int^{N_f}_{0} \tilde{m}(t) dt}{\tilde{s}}}
\end{equation}

 We can propose a reasonable guess for a solution of Eq.\ref{Eq9}. Let's start by assuming a solution of this form

 \begin{equation}\label{Eq12}
 \tilde{\Psi } (N) = e^{\int^{N_f}_{N_i} R(t) dt} = e^{\frac{\int^{N_f}_{N_i} \tilde{m}(t) dt}{\tilde{s}}}
 \end{equation}
 
 where R(N) is the OM-Knot curvature
 
 Essentially what we are saying is that a very reasonable guess  for the OM-QM  counterpart of the free particle quantum state is a topological invariant of the OM-Knot and concretely, the exponential of the total Knot curvature. This invariant is in general a function counting the statistics of possible crossing at the fuzzy center of the OM-Knot. From Eq.\ref{Eq12} we get 
 
 \begin{equation}\label{Eq13}
R(N) =  \frac{\tilde{m}(N) }{\tilde{s}}
\end{equation}

Using the presence of the total OM-Knot curvature in the formula for $  \tilde{\Psi } (N) $ we can find a more explicit and meaningful form of that solution. From the work of Mazur it is known that the crossing statistics of fuzzy Knots are given by logarithms of the Prime Numbers \cite{mazur2012primes} \cite{morishita2009analogies} \cite{morishita2011knots}. To each fuzzy Knot one can assign a prime and to each prime one can assign a fuzzy Knot. Mazur gave the following correspondence 

 \begin{equation}\label{Eq14}
 \log (\textit{p}) =  vol(K)  
 \end{equation}

where \textit{p} is a prime number and vol(K) is the volume of the hyperbolic complement of a Knot \cite{callahan1998hyperbolic}. When talking about a fuzzy Knot that prime \textit{p} counts the crossing statistics at the fuzzy center. By the other side that volume is equal to the total curvature integrated over the Knot. Therefore, we have 

 \begin{equation}\label{Eq15}
\int^{N_f}_{N_i} R(t) dt = \log (\textit{p}) 
\end{equation}

At this point we make contact with Number Theory by recognizing that in the formula above we can use the von Mangoldt function $ \Lambda (\textit{p}) $ instead of $ \log (\textit{p}) $. The von Mangoldt function is defined by 

 \begin{equation}\label{Eq16}
\Lambda (\textit{p}) = \bigl\{ \begin{array}{cll}
\log(\textit{p})& if & N = \textit{p} ^k  \\ \\
0 & &otherwise 
\end{array}
\end{equation}

from this we get the following possible solution for the OM-QM transformed Dirac Equation 

\begin{equation}\label{Eq17}
\tilde{\Psi } (N) = e^{(\sum_{q={\textit{p}^r}} \Lambda (\textit{N}) \delta (N-q))}
\end{equation}

Next, we found the OM-QM counterpart of Mass and  Energy

 \begin{equation}\label{Eq18}
\tilde{m}(N) = \tilde{s} \sum_{q={\textit{p}^r}} \Lambda (\textit{q}) \delta (N-q)
\end{equation}

 \begin{equation}\label{Eq19}
\frac{\tilde{E} ^2 }{(4 \pi^2)^2} = \tilde{s}^2 \sum_{q={\textit{p}^r}}   \sigma_q \log(q)q  ) \delta (N-q)  +\tilde{m} + 2 \tilde{m}^2 
\end{equation}

where sigma are the imaginary part of the zeroes of the Riemann zeta function. \\ \\

Finally, we want to emphasize that in all those results in [] we found a version of OM-QM which is a Number Theory based only in the variable N. That is, we only expressed OM-QM in terms of the scale of the OM-petal loops. We could do that because of the Observations were only geometrical in nature. In addition, the OM-Knot curvature in terms of that variable (that is, curvature from the OM-petals) accounts for all the global curvature we need for this kind of Observation. From now on we will depart from this and consider also observation of quantum states. That will force us to consider a separated curvature contribution coming from the n loops and their crossing in the central region.

\section{Quantum base states and mixed states in an Observation Modular framework }

As the central motive of the OM-correspondence is modularization by Observation we must recognize that the scope of observation in Step 3 is only partial. Step 3 refers only to modularization in space-time regions where Observation is implied. However in QM we not only observe positions in space-time but also quantum states. We start with a mixed state represented as a linear combination of certain base quantum states with certain weight coefficients. Those base states are eigenvectors of the operators representing quantum observables and span a Hilbert Space for the quantum system we are observing. A mixed quantum state in that base can be written as 

\begin{equation}\label{20}
|\Psi> = \sum_{i=1}^n C_i|\phi_i>
\end{equation}

The size n of the base of states is something we can get from QM and depends on the system we are analyzing. It depends on the Physics, geometry, topology and symmetries of the physical system at hand. QM give us explicit techniques for calculating those base states and their respective coefficients for forming mix states. The first question we want to answer here is: Where all that information is mapped in the OM-correspondence ? Or otherwise, what is the OM-correspondent of those base states and mixed states? In order to answer this we have to take a deep view on the way we constructed the OM-knot and Rosller-like structure representing the middle scale structure of OM-space. \\ \\ 

We start by noticing that there are two kind of loops in the OM-Knot and they are fundamentally different. From one side we have the so called OM-Knot "petals" originating from space-time modularization via Observation of position and time. From the other side we have the $ n $ loops coming from the topological freedom of winding around the central OM-Knot region. We will adopt the name contractible OM-Knot loops to refer to these later ones. Let's find first how many of those two kind of loops there are in the OM-Knot.\\ \\

Let's analyze in detail the difference between those two kind of OM-Knot loops. We start considering the OM-Knot "petals". In Figure.1 we see the number of possible space-time paths connecting initial and end states in a Feynman Amplitude is in fact infinite.  According to Step 3. in the OM-correspondence all those infinite paths are transformed in OM-Knot petals up to homotopy. That means that for example for the case of a free particle they will be mapped into exactly one OM-Knot petal as all the \textbf{infinite number of Feynman paths are homotopically equivalent in that case. Thus the number of OM-Knot petals is usually a small set and independent of the variable $ n $. Their number is equal to the number of geometric constraints in the space-time setup of the observed system. For example, each independent characteristic scale in the observed system will generate one OM-petal.}\\ 
There are some other characteristics of these OM-Knot petals which are important. Even when they can be deformed homotopycally, they cannot be contracted to the central region. There is a long,  almost parallel entry to the central region for both ends of such "petals" as it was discussed previously. This is a section of almost zero curvature in OM-space. We can  visualize that region as a thin, plane strip entering the central region. \\ 
How long is that entry region ? It depends on the metric distance between the end points in the source experimental setup in the QM side. we can assume that this will be a number of the form 

\begin{equation}\label{21}
l=\frac{\delta s }{\zeta_0}
\end{equation}

where $ \delta s $ is the length between the extremes of the Feynman paths. By the other hand, the Geometry of the system is also a component in the determination of the size of the quantum base of states. But other physical elements contribute too to the number $ n $. Those other elements are for example symmetries of the Hamiltonian or Lagrangian in the QM side. Also it is dependent on the observables we may choose for constructing the quantum base of states (that is on the concrete quantity we are observing or measuring). We may also choose a base of our Hilbert space made of a number of base states not totally linearly independent. However we see that the minimum number of linear independent base states is totally determined by Geometry, Symmetries and Physics of the system. \\ Thus, $ n $ can in general be an arbitrarily number larger or equal to  that minimum possible one. This exhibits  the footprint of its origin in the  topological winding freedom around the central region of the OM-Knot. \\ \\  

 The central, fuzzy region of the OM-Knot is the place where the information related to quantum states is mapped in the OM-correspondence. From the Rosller like structure of OM-space we see also that the OM-Knot petals are the parts of the Rosller flow which transit into the high dimensional part of this object. They are the part of the Rosller flow which escapes into the higher dimension and reinserts itself into the low dimensional flow. That low dimensional flow in the Rosller structure is precisely the one corresponding to the $ n $ contractible OM-Knot loops. They are the lower dimensional part of the Rosller flow. In holographic Theories as OM-QM that lower dimensional part of the space is the one pertaining to the Quantum content of the Theory. \\ \\
 
Let's now focus on the $ n $ contractible OM-Knot loops. A proper topological variable would not depend on the any scale coming from the geometry in the QM side. We know though that the size of the quantum base depends on the geometry of the observed system. Thus, n also does. It can also depend on its Symmetries.  \textbf{We propose the minimum number of those contractible OM-loops to be equal to the size of the Hilbert Space in the QM side. We can consider that the OM correspondents of the n quantum base states are n abstract vectors labeled by the $ n $ contractible OM-loops. One of the possible braids we can form from them will represent the OM version of a possible mixed state. The vector space spanned by these n base vectors is what we call OM-Hilbert space} \\ \\

Up to now it has been enough with defining the OM-knot as a fuzzy knot. We meant by this that the central region of the OM-Knot hides a combinatorial or statistical structure. That is, the fuzzy center represents the set of all the ways we can define the braid-like crossings of $ n $ loops which join at the minimal central cell in the OM-correspondence (or its equivalent Rosller-like structure). Into this central fuzzy region also do enter one or more OM-petals.  \\ 
In Figure 3 we present a visualization of this region including also one example OM-petal loop.  \\ \\

\begin{figure}[h]
	\centering
	\includegraphics[width=0.5\linewidth]{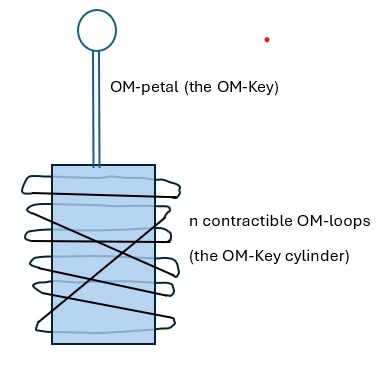}
	\caption[Figure 3]{OM-Knot central region visualization}
\end{figure}

 The number of external threads entering and leaving that braid-like region is the variable n. The number of possible braids we can form is $ (n)^2 $. We propose each of these braids correspond to a possible mixed state in the quantum side.  The visualization in Fig 2.b shows the difference in the way OM-petals and contractible OM-loops enter to this central region. OM-petals enter the central region in a planar way (curvature close to zero) but the threads in the OM-braid can enter this central region in different curvatures. They produce different crossings and therefore different sum of OM-Knot curvature. We will call this alternative visualization of the central region of the OM-Knot a Key-cylinder structure. We will use the term OM-Key cylinder structure because this region resembles an usual Key and key-cylinder mechanism where the OM-petals represent the Key and the OM-braid represent the Key cylinder.  By now, in Figure 3 we keep the Key and Key cylinder disconnected as this reflects the fuzzy, combinatorial nature of the OM-Knot center. We will see later that this OM-Key cylinder mechanism is indeed a perfect visualization of the modularization of quantum base states. In addition, the OM-key "turn" (high dimensional rotation) and OM-cylinder "opening" will be be a perfect representation of the mechanism to which the quantum state collapse gets mapped in OM-QM. The rest of this paper will be devote to the construction and consequences of this OM-Key cylinder region. \\ \\

 We will call the topological or "quantum" sector of OM-space an OM Hilbert space (or for brevity OM-H space). Mathematically, OM-H space has a vector space structure. \\ \\
 
 The full states space of OM-QM will be the space product of the "petals" sector and the OM-H space constructed above. We will call it OM-Riemann-Hilbert space (OM-RH space). We will show later in this work that this OM-RH space has more structure than a product vector space, it has the structure of a Riemann Surface. Along this work we will be focusing more on this full OM-RH space and not particularly on the OM-H or OM-petals sectors. \\ \\
 
 From the Rosller and OM-Key cylinder visualizations we recognize one more relation between the variables l and n. As shown in Figure 3 they are represented by dimensions which meet perpendicularly at the fuzzy center. We will represent this by defining a complex variable z in this way 
 
 \begin{equation}\label{22}
 z = u + i~ t
 \end{equation}
 
 where u represent the high dimensional sector, that is the scale variable for the OM-petals running from 0 to l. t represent the low dimensional sector, that is the OM-state index running from 1 to n. This will be the variables we will use in the rest of this work for talking about OM-RH in a unified way (or about is sectors). For example, when talking about high dimensional curvature we will be using the real part of z. We want to emphasize the fact that this complexification of the space is imposed on us by the very structure of OM-RH.\\ \\

\section{The solution to the free particle OM-Dirac Equation on OM-RH space}

Now let's calculate an explicit form of the solution to the OM Dirac Equation based on the OM-H base states labeling we have formulated before. Let's start by splitting the total OM-Knot volume in the this way 

\begin{equation}\label{23}
vol_{Tot} (l,n)= vol_R(l)+ i vol_H(l, n)
\end{equation}

\begin{equation}\label{24}
vol_R(l) = \int_{0}^{l} R_R (u) du 
\end{equation}

\begin{equation}\label{25}
vol_H(l, n) = \int_{0}^{l} R_H (n,t) dt 
\end{equation}

where $  R_R $ is the curvature coming from the OM-Knot petals sector and $ R_H $ the one coming from the OM-H sector. The i factor in front of $ vol_H $ comes from the perpendicularity of the OM-H and OM-R sectors of OM-RH.\\

Having now an statistical model of the central region of the OM-Knot in terms of the OM-base states we can write the volume $ vol_H $ as  

\begin{equation}\label{26}
vol_H(l,n) =  \sum_{j=1}^{n} \int_{0}^{l|_n}  {R_H}_j(t) dt
\end{equation}

where the $ R_j $ are the curvature of individual contractible OM-loops. This curvature term is completely new in OM-QM and reflects the dependency of the length of the OM-base of states on both the length of the QM base of states and also on the scale of the OM-petals. We can visualize the limits in this integral as defining a cut on the integration on each loop individually. It ties the curvature in the OM-H sector to the scale coming from the OM-R sector. This is a concrete realization of an analog of Quantum Gravity in OM-QM. We will see in next section that this is also a  footprint of the OM-Key cylinder mechanism in action.  \\ \\

Then we obtain the following total OM-Knot curvature 

\begin{equation}\label{27}
vol_{Tot}(l, n) = \int_{0}^{l} R_R (u) du +  i \sum_{j=1}^{n} \int_{0}^{l|_n}  {R_H}_j(t) dt 
\end{equation}

where we are using the components of the variable z to indicate that in reality these are complex integrals running in perpendicular directions ranges. Put otherwise, there is a hidden i sign in front of the integral for the low dimensional sector (the OM-H sector, $ R_H (t) $)

From this we get the following general form for the OM analog of a quantum mixed state 

\begin{equation}\label{28}
\tilde{\Psi}_{mix} = \exp{ (\int_{0}^{l} R_R (u) du + i \sum_{j=1}^{n} \int_{0}^{l|_n}  {R_H}_j(t) dt )}
\end{equation}

\section{Punctual measurement and state collapse in OM-RH space}

Any punctual measurement in QM has the following two characteristics

\begin{itemize}
	\item A unique geometrical scale $ l_1 $ is selected. This depends on the scale of the measurement apparatus or the setup of the observation.
	\item After the measurement the quantum state of the system collapses to exactly one of the quantum base states. This end base state is selected in a random way.
\end{itemize}

When we repeat the same measurement many times we get all the possible base states as the observation result with a certain probability distribution. For the i-th base state we have the probability $ p_i $ give by the Born rule 

\begin{equation}\label{29}
p_i = |c_i|^2 
\end{equation}

where the $ c_i $ are the coefficients of the i-th base state in the mixed state representing the quantum system.

Let's see what is the OM-QM correspondents of all those Physical side characteristics  of a measurement. \\ \\

In this section we will talk only about punctual measurements. \\ \\
Let's start with the first element in Eq.\ref{28} Let's recognize that fixing a scale $ l_1 $ (for example a metric distance) in the QM side translates to the introduction of a second scale different than the scale l we have in Eq.\ref{28}. Up to the moment of the measurement all the scales from 0 to l (or longer) were relevant for potential Feynman paths and therefore for the OM-Knot petals. However from the moment of the measurement of a concrete scale $ l_1 $ only scales in the range 0-$ l_1 $ will be relevant. The range of scales $ l_1 $-l (or longer) will be not relevant anymore. This introduces a cut of the OM-RH space to scale $ l_1 $. The effect of this cut in the mixed state in Eq.\ref{28} is a dependency only in $ l_1 $ and n. We will see later a visualization of this scale cut using the Rosller and OM-Key cylinder representations of the OM-RH space micro structure. We can reflect this in the OM-mixed state Eq.\ref{28} by splitting the total volume in the exponent in two parts, one from 0 to $ l_1 $ and one from $ l_1 $ to l.

\begin{equation}\label{30}
\begin{split}
\tilde{\Psi}_{mix} \downarrow_{l_1} = \exp{ (\int_{0}^{l_1|_n} R_R \downarrow_{l_1} (u) du + i \sum_{j=1}^{n} \int_{0}^{l_1|_n}  {R_H \downarrow_{l_1}}_j(t) dt )}* \\  \exp{ (\int_{l_1|_n}^{l} R_R \downarrow_{l_1} (u) du + i \sum_{j=1}^{n} \int_{l_1|_n}^{l}  {R_H \downarrow_{l_1}}_j(t) dt )}
\end{split}
\end{equation}

where the second exponential in that formula should be equal to 1. Thus

\begin{equation}\label{31}
\int_{l_1|_n}^{l} R_R \downarrow_{l_1} (u) du +  i \sum_{j=1}^{n} \int_{l_1|_n}^{l} {  (R_H \downarrow_{l_1}) }_j(t) dt = 0
\end{equation}

where $ \downarrow_{l_1} $ means the quantities where we have fixed l to $ l_1 $. This means we can ignore all the OM-contributions to the curvature (and OM state) above that scale $ l_1|_n $. Let's note that the total length of the OM-petal has then got reduced to $ l_1|_n $ . Thus the length of the almost flat region of the OM-key is given by 

\begin{equation}\label{32}
\lfloor (l_1|_n)/2 \rfloor
\end{equation}

where the $\lfloor (l_1|_n)/2 \rfloor $ means the integer approximating the number $ (l_1|_n)/2  $ from below. \\ \\

Let's now give an interpretation to Eq.\ref{31}. It means that the curvature of the OM-petals compensates the curvature of all the loops from scale $ l_1|_n $ to scale l. Looking at the OM-Key cylinder representation of the OM-RH space this means that all the contribution of contractible OM loops beyond scale $ \lfloor (l_1|_n)/2 \rfloor $ can be ignored (when we have set scale $ l_1 $ with an OM-H base of size n)

Thus we can write Eq.\ref{28} in this way 

\begin{equation}\label{33}
\tilde{\Psi}_{mix} \downarrow_{l_1} = \exp{ (\int_{0}^{l_1|_n} R_R \downarrow_{l_1} (u) du + i \sum_{j=1}^{\lfloor (l_1|_n)/2 \rfloor} \int_{0}^{l_1|_n}  {R_H \downarrow_{l_1}}_j(t) dt )}
\end{equation}

Therefore by fixing the scale in the QM side in a measurement we have limited the number of OM base states contributing the OM mixed state to $ \lfloor (l_1|_n)/2 \rfloor $. \\ \\

There exists an equivalence relationship between all those $ \lfloor (l_1|_n)/2 \rfloor $ base states which is not available for all the rest of states up to number n. It is related to the fact the OM-Key can reach them and operate on them according to the following procedure \\ \\ 

Any two of the base states from 0 to $ \lfloor (l_1|_n)/2 \rfloor $ are equivalent under the following process: we can pass from one state to the other one (in OR-RH space) via plane connections of petals to contractible loops followed (if there are crossings in the contractible loops) by rotations of the OM-petals (in the high dimensional sector) and final projection onto the OM-H sector (further insertion of the OM-Key). We will call this the OM-Key cylinder mechanism. The OM-petals represent the Keys and the OM-braid represent the Key cylinder. The OM-petals connection to OM-loops, rotation and projection represent the process of Key turning and key cylinder opening. In Figure 4 we show a visualization of this whole process

\newpage 

\begin{figure}[h]
	\centering
	\includegraphics[width=0.8\linewidth]{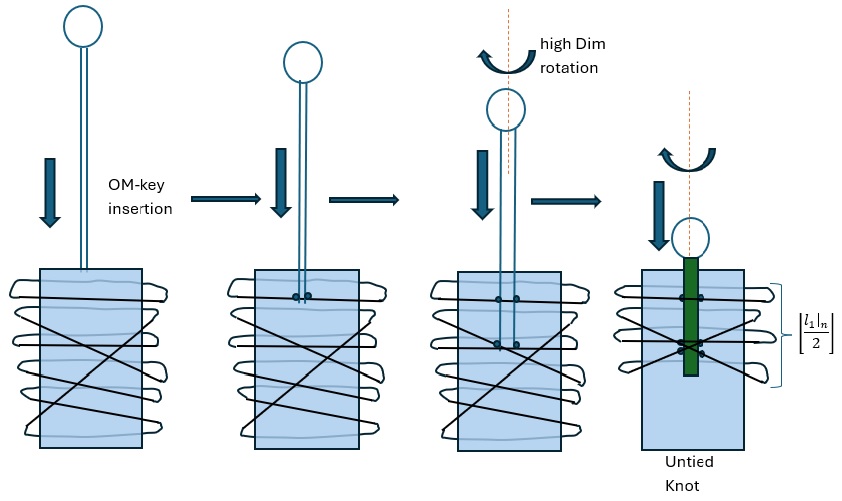}
	\caption[Figure 4]{The OM-Key cylinder mechanism}
\end{figure}

Regarding the second characteristic in a quantum measurement we notice the following: If there was an Observation of one quantum state then we need to extend the scope of the OM-correspondence to cover also this kind of Observations. We will add the following new Step 4 in the OM-Correspondence for the cases where we are considering also the Observation of quantum states.\\ \\

\textbf{OM-correspondence, Step 4 \\
If an observation of one unique quantum base state (from a base size of n) at scale $ l_1 $ has occurred then this is translated in OM-QM in the following reduction of the OM mixed state: we will identify (modularize) all OM base states contributing to the OM mixed state which are equivalent via the OM-Key cylinder opening process. We can ignore the other which are not equivalent via that mechanism. } \\ \\

Here we pause to mention something which is hidden in this new step: which concrete $ \lfloor (l_1|_n)/2 \rfloor $ OM-base states we are modularizing and which ones we are ignoring ? To give a precise answer we need to define some ordering in the OM base of states. This is not totally needed for the results in this paper but it is a question that comes automatically when reading Step 4. Fortunately there is an natural ordering for these OM-base states. As each one is labeled by one concrete integer from 1 to n, and each one of those integers can be mapped to the same state index in the quantum side, we can assume that this n is associated to an increasing Energy scale in the quantum side. As we saw in \cite{frugone2023quantum} and in Section 1, Energy is mapped in OM-QM into certain functions of the zeroes of the Riemann Zeta Function. Thus, we can assume that the index in the OM-base states represent increasing $ \tilde{Energy} $ levels, that is they are associated to increasing zeroes of the Riemann Zeta Function. We won't need this ordering for this work but this gives a more complete context to Step 4. 

This new Step in the OM-correspondence is the quantum analog of Step 3. It extends OM-QM to cover observation of quantum states with a simultaneous description of an analog mechanism for the unavoidable state collapse which is occurring in the QM side. We will show that if we apply Step 4. to  Eq.\ref{33}  we will get a (dynamically) reduced state equal to exactly one unique OM-base state. \\ \\
Let's make use of the visualization of the OM-Key opening mechanism and how it acts in Eq.\ref{33}  

\newpage

\begin{figure}[h]
	\centering
	\includegraphics[width=0.6\linewidth]{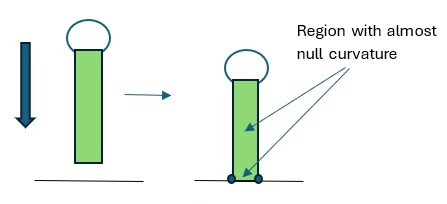}
	\caption[Figure 5]{OM-Key "opening" contractible OM loops with no crossings}
\end{figure}

\begin{figure}[h]
	\centering
	\includegraphics[width=0.6\linewidth]{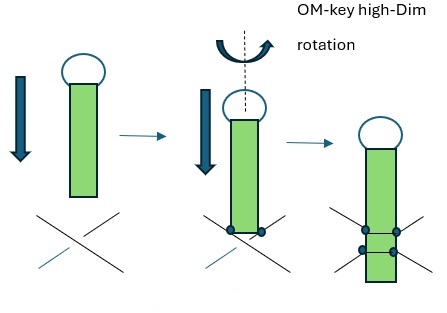}
	\caption[Figure 6]{OM-Key "opening" contractible OM loops with crossings}
\end{figure}

The OM-Key insertion into the OM-key cylinder occurs via connection of the two almost parallel sections of the OM-Key to the successive levels of the OM-braid it meets in way down the OM-cylinder. The OM-Key enters in the OM-key cylinder close the OM-Knot central region. If there is no crossing in the braid at this level, then the braid just continues its insertion by connecting to the next level of the braid. At the same time the OM-Key consumes one of its levels (2 elements of the scale $ l_1 $ total length). If in the next level it meets a braid crossing then in order to continue its insertion the OM-Key will need to perform a half full turn in high dimension sector. This OM- Key rotation has a + or - sign according to the crossing sign. In case we meet in this level a braid crossing with two crossings of the same threads then the OM-Key rotates twice in the same direction and the net effect is returning to its initial orientation. That is the same as a zero rotation. The same will occur for any even number of crossings in the OM-braid at a certain level. If we consider the braid representing a certain product of states in the OM-H space then that property can be rephrased as: any square of OM-base states can be identified with a zero Key rotation. Any non square product of two OM-base states will be either associated to a +1 or -1 rotation of the OM-key according to the pair or impair number of crossings it contains. We notice that there is a well known Number Theory function whose behavior is similar to the the previously described  action of the OM-Key on the OM-key cylinder. It is the Moebius function defined by 

\begin{equation}\label{34}
\mu (N) = \left \{ \begin{array}{rcl}
0 & if & non-square-free \\
+1 & if & \mbox{square-free~and~even~number~of~prime~elements} \\
-1 & if & \mbox{square-free~and~odd~number~of~prime~elements} \\
\end{array}\right.
\end{equation} \\ \\

Let's analyze the effect of this OM-Key insertion in the OM-mixed state

\begin{equation}\label{35}
\tilde{\Psi}_{mix} \downarrow_{l_1} = \exp{ (\int_{0}^{l_1|_n} R_R \downarrow_{l_1} (u) du + i \sum_{j=1}^{\lfloor (l_1|_n)/2 \rfloor} \int_{0}^{l_1|_n} {R_H \downarrow_{l_1}}_j(t) dt )}
\end{equation}

After the full insertion of the OM-Key in all its length the first term in the exponent gets reduced to 0. The OM-Key has been consumed in its totality and left with only one element of minimal size. Thus we are left only with the second term in the exponent. As discussed above, the OM-key cylinder opening mechanism can be modeled by the action of the OM-Key on this second term. We propose this action is well represented by the following 

\begin{equation}\label{36}
\tilde{\Psi}_{col} \downarrow_{l_1} =  \exp(i ~\tilde{\alpha} \sum_{j=1}^{\lfloor (l_1|_n)/2 \rfloor} \int_{0}^{l_1|_n} \mu(t) dt)
\end{equation}

where we have collapsed all the sub volumes of all the $ R_j $ to a unit minimal volume $\tilde{\alpha}$. We will call this an OM-collapsed state and no longer a mix state. By summing under the integral sign technique we get the following

\begin{equation}\label{37}
\tilde{\Psi}_{col} \downarrow_{l_1} =  \exp(i~\tilde{\alpha}   \sum_{j=1}^{\lfloor (l_1|_n)/2 \rfloor} \mu(j) \int_{0}^{\lfloor (l_1|_n)/2 \rfloor} dt)
\end{equation}

A well known property of the Moebius function is that its sum from 1 to k over all the divisors of k is 1. 

\begin{equation}\label{38}
\sum_{j|k}^{k} \mu(j) = 1
\end{equation}

and as we imposed in our Step.4, the action of the OK-Key opening mechanism is simulated by the factor $ \mu(j) $. After the $ \lfloor (l_1|_n)/2 \rfloor $ rotations of the key the OM-knot should be totally untied. There should remain only one contractible loop. That means that 

\begin{equation}\label{39}
\sum_{j=1}^{\lfloor (l_1|_n)/2 \rfloor} \mu(j) = 1
\end{equation}

and then we get the final form of the collapsed state

\begin{equation}\label{40}
\tilde{\Psi}_{col} \downarrow_{l_1} =  \exp(i~\tilde{\alpha} \lfloor (l_1|_n)/2 \rfloor )
\end{equation}

This is exactly the OM-base state number $ \lfloor (l_1|_n)/2 \rfloor $. The full OM-mixed state (from a OM-base of size n) after fixing the state to $ l_1 $ and applying the OM-correspondence with its four steps collapses to exactly one OM-base state, and concretely the $ \lfloor (l_1|_n)/2 \rfloor $ one. \\
This is not a totally random collapse as we would expect in the QM side but it is an almost random one. This can be seen easily as the number $ l_1 $ is usually a huge number (as it is a quotient of a distance by the Plank length) and n can be in general a rather small number. In some experimental setups where we would be for example measuring spins of an electron this number n can be as low as 2. The number $ \lfloor (l_1|_n)/2 \rfloor $ for such different values of $ l_1 $ and n is a almost chaotic under small variations of the scale $ l_1 $. For n=2 one gets exactly $ 50\% $ of the time a 0 and  $ 50\% $ of the time a 1. This reproduces exactly the random behavior in the QM side also in this collapsed OM-state. What is new here is that we have a concrete form for this pseudo randomness and we see it is not totally random for some different relationships of scale between $ l_1 $ and the size of the quantum base of states. Thus we argue that the full OM-correspondence (including Step 4) is an explicit realization in OM-QM of all the elements of the Quantum state collapse in the QM side. However the OM-QM mechanism differs from the QM one in which for a given $ l_1 $ and n there is no randomness at all. The selected collapsed base state is completely deterministic and given by $ \lfloor (l_1|_n)/2 \rfloor $.  Small variations of $ l_1 $ provide an pseudo chaotic behavior of the resulting collapsed state. Thus, in realistic physical scales and quantum base sizes in the QM side, the OM-side state collapse mechanism looks almost random. \\ \\
As we constructed OM-QM using a sort of covariance between the form of equations in the QM side and in the Mathematical side, we propose that also this pseudo-chaotic OM-state collapse mechanism may be saying something about the Physical side. We propose this pseudo randomness of the quantum state collapse is something also present in the Physical side. The concrete law governing the collapse is totally deterministic and also valid for the physical side ($ \lfloor (l_1|_n)/2 \rfloor $) and only looks random in QM for some ranges of n and $ l_1 $ but not for others. We will analyze later those ranges where we could see some deviations from total randomness in the quantum state collapse. This constitutes an experimental prediction of OM-QM which could serve for falsifying it. It is not strange that dealing with a concrete mechanism for the state collapse we end up in something which seems to be an extension of QM. After all there is no model of this process in QM itself. Any modeling of this process (even coming from OM-QM) nECCssarily means an extension of QM itself.

\section{Repeated measurement in OM-RH space, the OM-Born rule}

We focus now in the last characteristic of measurements, the Born rule for the result of repeated measurement. To find the OM-correspondent of the Born rule we need to understand how the coefficients of a quantum state (on the base of its Hilbert space) are mapped in OM-QM (on the base of OM-base states). In QM those coefficients are given by the inner product of the mixed state vector on each of the base states. The Born rule states that the square of the module of such coefficients represent the probability of getting the corresponding base state in repeated experimental measurements of the system. \\ \\
 Up to now we have focused on a more topological view of the mixed state Eq.\ref{33} where the most important was visualization of the global aspects of the OM-Knot and its volumes. Now we will use the following expansion in terms of the OM-base states

\begin{equation}\label{41}
|\tilde{\Psi}>_{mix} = \sum_{j=1}^n \tilde{C}_j|\tilde{\phi}_j>
\end{equation}

where the $ |\tilde{\phi}_j> $ are the n base vectors spanning the OM-H space. Each j is a label representing a unique contractible OM-loop. The coefficients $ \tilde{C}_j $ are the analogs of the Hilbert space expansion coefficients in OM-QM. As we saw in previous section, we can write this expansion as the real number of really non OM-equivalent braids we can form from the n non-contractible loops is exactly n. \\
In previous section we found that for a given scale of observation  $ l_1 $ and a quantum base size n the result a punctual measurement is not probabilistic at all, it always falls in a deterministic way to the base state $ \lfloor (l_1|_n)/2 \rfloor $ . If we assume that the number n of total possible OM base states is fixed then any statistical variation in the resulting state on repeated measurements must come only by uncertainty in the scale $ l_1 $ . That uncertainty in $ l_1 $ is given by the Heisenberg relation which in the OM-QM becomes 

\begin{equation}\label{42}
\Delta \tilde{l_1} = \frac{2 \pi i}{\Delta \tilde{P_1}}
\end{equation}

where $ \Delta \tilde{P_1} $ is the OM-correspondent of an uncertainty in momentum. As we are in the conditions of the Central Limit Theorem we know the spread of the states around to the center state  $ \lfloor (l_1|_n)/2 \rfloor $ is given by a Normal Probability Distribution as shown in this figure \\ \\

\begin{figure}[h]
	\centering
	\includegraphics[width=0.5\linewidth]{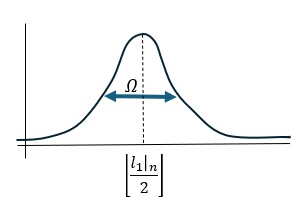}
	\caption[Figure 7]{probability of getting another OM-state when measuring at scale $ l_1 $ with an OM base of size n}
\end{figure}

The concrete probability formula takes the following form

\begin{equation}\label{43}
P(l_1, n, \lfloor t \rfloor) = \frac{\exp(-\frac{(\lfloor t \rfloor -\lfloor (l_1|_n)/2 \rfloor)^2}{2 \pi \Omega^2})}{\sqrt{2 \pi \Omega}}
\end{equation}

where $ \Omega $ is the middle width of the distribution given by

\begin{equation}\label{44}
\Omega(l_1,n) = \lfloor \frac{(l_1+ \frac{2 \pi i}{\Delta \tilde{P_1}})|_n}{2} \rfloor - \lfloor \frac{(l_1- \frac{2 \pi i}{\Delta \tilde{P_1}})|_n}{2} \rfloor
\end{equation}

Thus we propose that those$  P(l_1, n, \lfloor t \rfloor) $ are the equivalent of the probability for the OM-base state $ \lfloor t \rfloor $ to appear after a punctual measurement where the state $ \lfloor (l_1|_n)/2 \rfloor $ was observed. That is a conditional probability. In OM-QM in a OM-RH space we are always past the point of one punctual measurement. Thus we are also past the point of a fixing of the scale $ l_1 $. And this scale also determines the state $ \lfloor (l_1|_n)/2 \rfloor $. Then in OM-QM the  Born rule will be talking not about the probability of one concrete OM-base state but instead about the conditional probability of observing other OM-base states having already observed $ \lfloor (l_1|_n)/2 \rfloor $. In the QM side we repeat the same measurement with exactly the same geometrical parameters, that is every time we are measuring at a scale spread around $ l_1 $. We propose the square root of that probability distribution Eq.\ref{43}  are the OM-QM version of the QM coefficients of a quantum mixed state. Thus we can rewrite the mixed state in this way

\begin{equation}\label{45}
|\tilde{\Psi}>_{mix} = \sum_{\lfloor t \rfloor = 1}^n \sqrt{P(l_1, n, \lfloor t \rfloor)} ~ |\tilde{\phi}_{\lfloor t \rfloor}>
\end{equation}

The OM-coefficients given in this way do depend on the parameter $ \Delta \tilde{l_1} $. We don't need to get an explicit form to that parameter. For us it is enough to understand the origin of each term coming from the QM side. We don't intend getting a totally explicit form of those coefficients at this point. We also notice that the OM-coefficients of the OM-mixed state are dependent on the scale $ l_1 $ of the repeated measurement . In QM the coefficients are also partially dependent on the scale of the repeated measurement. However in OM-QM there is a completely deterministic connection between $ l_1 $ and the OM-base state which we can obtain after a punctual measurement. Such connection does not exist in QM, not in such an explicit way. And this comes from the fact that in OM-QM we are always past the observation at a scale  $ l_1 $ spread via the OM-Heisenberg relation.

\section{An analytical model for the OM-state collapse mechanism}

In  Section 5 we saw that the OM-Key cylinder mechanism can be visualized as a consumption of the OM-petal (the Key) by the contractible OM-loops (the key cylinder) that give us as a result a unique OM-base state. Thus this process changes a OM-petal into a single contractible OM-loop. We will see next that OM-QM give us a analytical mechanism by which this OM-state collapse can be described. With this we will get a more operational form of Step 4 of the OM-correspondence which will allow us to apply it to diverse measurement setups. There is no analog in pure QM for this detailed dynamical view of the OM-state collapse. We will see that OM-QM can guide us in that uncharted territory and that we will be able to give some Physical interpretation to that detailed collapse mechanism. \\ \\ 
We will start by proving that there is an analytical way to pass from an state represented only by an OM-petal to an unique contractible OM-loop. We start with an OM-state for one OM-petal where we assume an observation at scale $ l_1 $. 

\begin{equation}\label{46}
\tilde{\Psi}_{mix} \downarrow_{l_1} = \exp{ (i~\int_{0}^{l_1|_n} R_R (u) du ~)}
\end{equation}

As discussed previously ignoring the OM-H sector means we are only observing here the space-time geometry properties of the system (for example a metric distance or position). We know that a possible solution for $ R_R $ is simply 

\begin{equation}\label{47}
R_R (u) = \tilde{\alpha} \frac{1}{u} 
\end{equation}

Let's make the following transformation of variables:

\begin{equation}\label{48}
u \longleftrightarrow \zeta(t)
\end{equation}

that is, we are stretching the coordinate u to the Riemann zeta function of the t coordinate. In addition to a dilation it contains a rotation of 90 degrees in the complex plane of the variable z. After this transformation for $ R(u) = 1/u $ we get

\begin{equation}\label{49}
\tilde{\Psi} \downarrow_{l_1} =  \exp(i~\tilde{\alpha}  \int_{0}^{\zeta^{-1}(\lfloor (l_1|_n)/2 \rfloor)} \frac{\zeta'(t)}{\zeta(t)} dt~)
\end{equation}

We substitute in this state the so called Riemann-von Mangoldt relationship from Analytical Number Theory 

\begin{equation}\label{50}
\frac{\zeta'(t)}{\zeta(t)} = - \sum_{q=1}^{\infty} \frac{\Lambda(q)}{q^t}
\end{equation} 

where $ \Lambda(q) $ is the von-Mangoldt function. We get 

\begin{equation}\label{51}
R(t) =  - \tilde{\alpha}  \int_{0}^{\zeta^{-1}(\lfloor (l_1|_n)/2 \rfloor)} \sum_{q=1}^{\infty} \frac{\Lambda(q)}{q^t} dt
\end{equation}

Exchanging the sum and integral sign this becomes 

\begin{equation}\label{52}
R(t) =  - \tilde{\alpha}  \sum_{q=1}^{\infty} \Lambda(q) \int_{0}^{\zeta^{-1}(\lfloor (l_1|_n)/2 \rfloor)} \frac{1}{q^t} dt = - \tilde{\alpha}  \sum_{q=1}^{\infty} \frac{\Lambda(q)}{\ln(q) (q^{\zeta^{-1}(\lfloor (l_1|_n)/2 \rfloor)} - 1)}  
\end{equation}

Due to the possible values of the von-Mandgoldt function we get the following 

\begin{equation}\label{53}
R(t) =   - \tilde{\alpha}  \sum_{q=p^k}^{} \frac{1}{q^{\zeta^{-1}(\lfloor (l_1|_n)/2 \rfloor)}} = - \tilde{\alpha}~ \zeta (\zeta^{-1}(\lfloor (l_1|_n)/2 \rfloor))
\end{equation}

where the previous sum runs over all powers of the prime numbers. We finally get  

\begin{equation}\label{54}
R(t) =   - \tilde{\alpha}~ \lfloor (l_1|_n)/2 \rfloor
\end{equation}

which is precisely the same state we got in Eq.\ref{40} for the collapsed OM-state after Step 4 in the OM-correspondence. We conclude that the transformation $ u \longleftrightarrow \zeta(t) $ is equivalent to applying that Step 4 of the OM-correspondence (having previously fixed the scale of the observation to $ l_1 $). This gives a very operational definition to Step 4. of the OM-correspondence. The complete OM-Measurement Process gets mapped into the following steps according the the OM-correspondence 

\begin{enumerate}
	\item An observation at scale $ l_1 $ occurs in a system with a quantum base of size n \\
	\item We apply Step 3. of OM-correspondence, (that is modularize space-time, or otherwise, we pass to OM-space)\\
	\item We stretch OM-space via $ u \longleftrightarrow \zeta(t)$ \\
	\item the OM-state gets mapped under that stretching into the OM-base state $ \lfloor (l_1|_n)/2 \rfloor $ state
\end{enumerate}

as we proposed before we conjecture that the last item can also be reflected in the Physical side. That is, the end result of the measurement is the quantum base state $ \lfloor (l_1|_n)/2 \rfloor $.

We conclude that the stretching $ u \longleftrightarrow \zeta(t)$ plays a very exceptional role towards OM-QM. It connects (in a very non trivial way) the high dimensional sector of OM-space with the low dimensional OM-H space. It looks like a very fine tuned projection of the high dimensional sector of OM-RH space into lower dimensional OM-H space.The efficacy of its action on OM-RH space comes from the Riemann-von Mandgoldt relation interplay with the OM-Dirac equation. 

\section{OM-QM dynamics on OM-RH space}

As we saw previously once we start considering the OM-correspondence with both Step 3 and Step 4 simultaneously we need to work on the complexified space $ z=(u+it) $. We also notice that the global topology of the OM-Key cylinder visualization of the OM-Knot center is the one of a Torus. The OM-petals represent the "large or slow" periodic variables of this Torus and the contractible OM-loops represent the "small or fast" periodic variables. More generally, the number k of OM-petals (or geometric constraints)corresponds to the number of holes in the resulting k-genus Torus. 

\begin{figure}[h]
	\centering
	\includegraphics[width=0.7\linewidth]{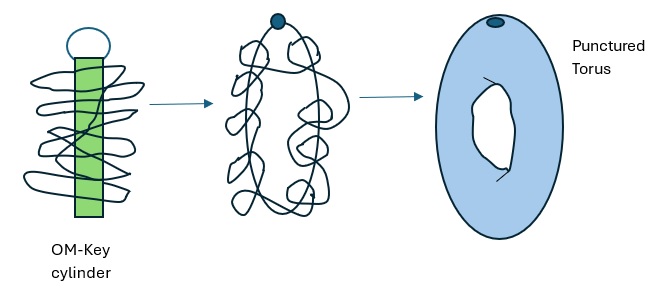}
	\caption[Figure 8]{OM-RH structure as a Riemann Surface}
\end{figure}

The periodicity in the Torus induces a modularization on meromorphic functions living on the z-complex space. That is, the only meromorphic functions living on that space which respect the global bi-periodicity implicit in OM-RH space are the ones which are also biperiodic on the z-space. This double periodicity in OM-RH space produces a tessellation of regions of z-space where those holomorphic functions have the same values. The Mathematics we are falling into here is the one of Riemann Surfaces \cite{teleman2003riemann}. OM-RH space is indeed a Riemann Surface. The regions in OM-RH space that we identify in Step 3 and Step 4 of the OM-correspondence correspond to lattices in the z-complex space. From the theory of Riemann Surfaces we know that the only meromorphic functions which have such bi-periodicity as those lattices are generated by the Weirstrass Phi Function. 

\begin{equation}\label{55}
\Phi(z) = \frac{1}{z^2}+\sum_{j=1}^{\infty} (2j+1)~G_{2j+2} z^{2j}
\end{equation}

where the functions G are the so called Einsenstein series 

\begin{equation}\label{56}
G_j = \sum_{0 \neq \lambda \in \Lambda} \lambda^{-j}
\end{equation}

are sums over the points belonging to the borders of the previously mentioned periodic lattice on z-space (excluding the origin of the z-plane). In an OM-QM context these G functions are contractions or inversions of the borders of the modular regions in OM-RH space. The only poles of such function occur on the borders of those lattices and they are poles of order 2. We propose that the Weirstrass function is indeed the general form of a mixed OM-state in the full OM-RH space.The reason for this is that OM-state must be bi-periodic in the "fast" u variable and "slow" t variable. In fact this is the mathematical meaning of taking Step 3 and Step 4 together. In addition we want the OM-state to be represented by a "well behaved" function on the z-complex space, that is a holomorphic function except for some eventual poles on the border of the module regions.  \\ \\

The Weirstrass $ \Phi  $ function respects the following differential equation 

\begin{equation}\label{57}
\Phi ' = 4 \Phi^3 - 2 g_2 \Phi - g_3
\end{equation}

where 

\begin{equation}\label{58}
g_2 = 60 G_4
\end{equation}

and 

\begin{equation}\label{59}
g_3 = 140 G_6
\end{equation}

are called the modular invariant. In OM-QM terms $ g_2 $ carries the meaning of the Key-cylinder mechanism where the OM-H cylinder is not having a crossing of contractible OM-loops, a four points function. $ g_3 $ in turns, has the meaning of the Key-cylinder mechanism where the OM-H cylinder has a crossing of contractible OM-loops, a 6 points function. \\ 

Eq.\ref{57} describes the dynamics of the state Eq.\ref{55} in OM-RH space. It is an explicit analytical realization of the evolution of the OM-mixed state after observation. It is a realization of the OM-Key cylinder mechanism between the points of start and collapse to an unique OM-base state. \\ \\

The invariants $ g_2 $ and $ g_3 $ do control the evolution. We can get a development of those two functions in terms of the OM-base states in the following way 

\begin{equation}\label{60}
g_2 = \frac{4}{3} \pi^4 [1 + 240 \sum_{k=1}^{\infty} \sigma_3(k) \exp^{2k (\pi i l_1 |_n)}]
\end{equation}

\begin{equation}\label{61}
g_3 = \frac{8}{27} \pi^6 [1 - 504 \sum_{k=1}^{\infty} \sigma_5(k) \exp^{2k (\pi i l_1 |_n)}]
\end{equation}

where 

\begin{equation}\label{62}
\sigma_a(k) = \sum_{d|k} d^a
\end{equation}

is the so called divisor function of k.  \\ \\ 

A Fourier Series form of the Weirstrass function  Eq.\ref{55}  can be written in this form 

\begin{equation}\label{63}
\Phi(z) = \frac{1}{z^2}+ (A+B) + z^2~\sum_{k=1}^{\infty} (240 A \sigma_3(k) -504 z^2 B \sigma_5(k)) \exp^{2k (\pi i l_1 |_n)}
\end{equation}

where 

\begin{equation}\label{64}
A = \frac{3.4.\pi^4}{3.60}
\end{equation}

and 

\begin{equation}\label{65}
B = \frac{5.8.\pi^6}{27.140}
\end{equation}

It is easy to prove that if we perform the substitution  $u  \longrightarrow  \zeta(t)$ in this Fourier series we get exactly the $\lfloor (l_1 |_n)/2 \rfloor$ OM-base state. For this to be true it enough proving that the sum in Eq.\ref{63}  fulfills the following relation 

\begin{equation}\label{66}
(240 A \sigma_3(k) -504 \zeta^2(k) B \sigma_5(k)) = \frac{\delta (2 k \pi - \beta)}{\zeta^2(k)}
\end{equation}

where $ \beta $ is an integer larger than $ 2 \pi $. Although this looks like a hard constraint to be fulfilled, it is valid for every value of $\beta$ we select. One particular check of  this validity is simply choosing the first $\beta$ possible which is equal to 1. Summing on k on both sides of Eq.\ref{66} and remembering that $\zeta(1) = \gamma$ (where $ \gamma $ is the Euler-Mascheroni constant) one gets 

\begin{equation}\label{67}
4.240 A -504 B = 8
\end{equation}

And substituting the values of A and B we can prove this identity is true. \\  We can consider Eq.\ref{66} as a Fourier transform equation for the collapse of the approximate OM-mixed state.The l.h.s of this equation is an explanation of the OM-Key cylinder mechanism. \\ \\
 
 Finally, we can compare Eq.\ref{63} to Eq.\ref{45} from Section 6 . They represent the same OM-mixed state (corresponding to a free particle in the QM side). In Eq.\ref{45} the coefficients of the development in that development had still several components which were left as variables without an explicit mathematical expression. Nevertheless, they showed explicitly their corresponding quantum side physical correspondent quantities. In Eq.\ref{63} there is no such unknown parts of the coefficients in the development. We cannot directly equate coefficients in both equations as in Eq.\ref{45} we have a development over the OM-base states and in the Weirstrass function we have in addition two first terms representing the order 2 pole and an order zero term (A+B). However, we can get an approximate form of the explicit form of Eq.\ref{45} if we consider it to be a Fourier Transform instead of a sum over a finite number of OM-base states. It would be the Fourier transform of coefficients having a Gaussian form. Thus, that Fourier transform is easy to calculate and also has a Gaussian form with an additional $ 1/z^2 $ factor. Then if we produce the Maclaurin development of this Gaussian explicit form of Eq.\ref{45} around  $ \lfloor (l_1|_n)/2 \rfloor $ we get an explicit form of Eq.\ref{45} which can be compared order by order with Eq.\ref{63}. The comparison of the coefficients in both development of orders \{-2,0,2,4,6\} in the variable z gives the following results: \\ \\
 
 \begin{itemize}
 	\item The pole of order 2 is reproduced in both sides \\
 	\item The order 0 coeffients gives the following expression for the width $\Omega$ in Eq.....
 	\begin{equation}\label{68}
 	\Omega = \frac{(A+B)^{\frac{1}{2}}}{\pi \sqrt{2}}
 	\end{equation} \\
 	\item We get the following explicit form for the coefficients in the development in Eq.... (for a the OM-correspondent of a free particle)
 	\begin{equation}\label{69}
 	C(l_1, n, k)=\sqrt{P(l_1, n, k)} = \exp[ -~\frac{( \pi^2 (k-\lfloor (l_1|_n)/2 \rfloor)^2)}{(A+B)}]
 	\end{equation}	
 \end{itemize}
 
 Let's keep in mind that this is valid only for the OM-correspondent of the quantum free particle. For quantum non-free systems one would get different probability distributions and then different coefficients. This is a topic which we intend to attack in future works.

\section{Entanglement and the EPR experiment in OM-RH space}

In order to have a complete mapping of QM into OM-QM we need to know what is the OM-correspondent of Quantum Entanglement. To explore this we will make use of a quantum setup where entanglement is present and where observation is also clearly defined both in space-time and in quantum states. This toy model is the EPR experiment which we describe in the following figure 

\begin{figure}[h]
	\centering
	\includegraphics[width=0.5\linewidth]{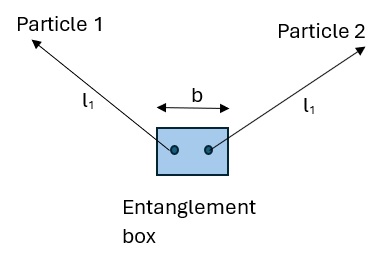}
	\caption[Figure 9]{setup of the EPR experiment}
\end{figure}  

First thing we can recognize is that we have three clear scales or geometric constraints in this setup. One assigned to each one of the two EPR particles and one third one assigned to both of them. This last constraint accounts for the fact that both EPR particles are initially entangled with each other in a region that we will call the entanglement box. Let's call b the scale of this box in units of the Plank Length.  \\ 

In the EPR experiment we measure the spin of one of the particles at a certain scale $ l_1 $ and we get one possible value of spin (in QM which of the possible values we get is  totally random). In the EPR experiment there is an aspect which is not usually discussed as in standard QM is not very relevant  to the Physics of the process. That aspect is that there at three observations involved in the setup. This is evident by the fact we have three space time constraints and these need to be observed. In OM-QM this is an important aspect of this experiment as OM-QM is all about modularizing spaces by regions where observation is implied. It is also important to notice that those three observations can in principle be done by different observers who later share their observations in order to get a consensus version of the Physics of the process. When we talk about observers here we can be referring to measurement apparatuses or the final observers, the experimenter who is interpreting the results of the measurements.  There could be one version of the experiment where only one final observer is needed but doing at least two observations. In this version  measurement devices record the values of spin at the same time and then that single final observer reads those values and forms a consensus view of the process. In EPR the experimental setup assumes conservation of all the angular momentum observables including spin. Then observing one of the possible values of  spin in one particle and measurement of the existence of entanglement in the entanglement box by one final observer is enough for knowing automatically the result of the recording by the second measuring apparatus. The deduction (and read out of the second measuring apparatus record) allows to this single final observer to form the consensus view of the EPR process. According to the Copenhagen interpretation of QM, until this final observer does not measure at least two of those observations nothing can be said about the possible result of the second measuring apparatus. Moreover, from an experimental point of view if we want to corroborate that the QM prediction for the values we will get in the second apparatus is true this cannot be done until we read the registered result in the second measurement apparatus. Thus, the formation of a real consensus view of the EPR process by this single final observer will need the three measurements. The EPR experiment and in general QM is all about the formation of this consensus view by one or many final observers. \\ 
There is another version of the EPR experiment where we have two final observers reading the results for spins registered at the same time by the two different final observers. At least one of those final observers has also observed the existence of the entanglement box. The consensus view of the EPR experiment will be formed when both final observers share their private or local observations. \textbf{According to the Copenhagen interpretation of QM there is no consensus view until this moment. In this version of EPR, up to this moment only one of the final final observers knows of the entanglement box for both particles and of all the conservation laws which apply to the common system. And that is all what is known until the moment of formation of the consensus view among observers. This aspect of EPR is crucial for OM-QM and we will see it translates in a concrete interpretation for Entanglement and the EPR experimental results.} \\ \\

Now let's see how this quantum side setup maps into OM-QM. As we have three geometrical constraints we know the OM-Knot will have three OM-petal loops. All of them have to keep the footprint of the spatial scales involved in them. Two of the petals will come from the modularization of the two $ l_1 $ scales. These two OM-petals will start and end at each of their respective particle minimum cells. The third OM-petal will have the same scale as the entanglement box. This third OM-petal connects both minimal cells with an added curvature which represents Entanglement. There will be at least one crossing between two of such loops in order to generate this extra curvature inside the entanglement box. By the other hand, the OM-H base will have two contractible OM-loops (corresponding to the two spin states in the quantum side). In the following figure we represent this geometry

\begin{figure}[h]
	\centering
	\includegraphics[width=0.5\linewidth]{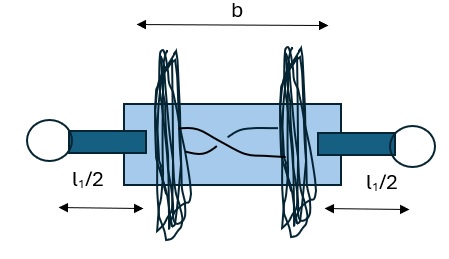}
	\caption[Figure 10]{Visualization of the OM-analog of entanglement in the EPR experiment}
\end{figure} 

There are two OM-Keys and two OM-Key cylinders in the EPR setup. Here we will make a crucial observation about the way these two OM-Keys and OM-Key cylinders are related to each other in their interaction region. This is described in more detail in the following figure 

\begin{figure}[h]
	\centering
	\includegraphics[width=0.5\linewidth]{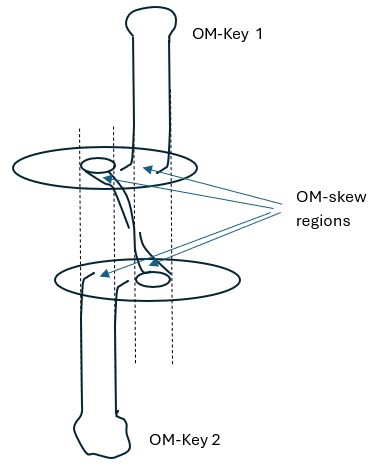}
	\caption[Figure 11]{The OM-skew regions of the OM-Knot. The joining regions of the high dimensional sector to the low dimensional one}
\end{figure} 

In the Rosller visualization of the way a OM-petal joins the low dimensional contractible OM-loops region we see that far from the joining region they come in a perpendicular direction (from a high dimension). However immediately close to the joining region they become almost parallel to the low dimensional OM-loops. This region is small as it happens all inside the central OM-Knot section, that is inside the minimal central cell of OM-space. Due to this small skew region we see in Figure 10 that the entry of the OM-Key to this joining region is always shifted respect to the central region of the contractible OM-loops. The center of the n contractible loops is shifted respect to the perpendicular line the OM-petal is entering. This s a typical characteristic of any Rosller attractor like object. It s a characteristic of how a high dimensional part of an object joins the low dimensional part. Every joining region of a OM-petal and the contractible OM-loops will have this skey region. Let's call this small joining regions the OM-skew regions of the OM-Knot.  \\ \\
In the EPR case we have four OM-skew regions as described in Figure 10. There are two mirror OM-skew regions one for each particle due to the long OM-petals. In addition we have two skew regions in the joining of the two entanglement loops between the two particles. The final result of those four OM-skew regions is that the OM-Key of particle 1 comes in a perpendicular direction directly over the central region of the n contractible loops of particle 2. A mirror geometrical situation happens in for the other particle. The OM-Key of particle 2 comes in perpendicular direction to the center of the n contractible loops of particle 1. \\ 
The OM-Key cylinder mechanism implies the OM-Keys will insert themselves in the central region following the perpendicular regions they are coming from. They will hit first the the crossing in the OM-entanglement box and will start "opening" the OM-cylinder from this point. However due to the series of OM-skews described above we see that this will result in OM-Key 1 opening the OM-Key cylinder of particle 2. Viceversa, the OM-Key 2 will open the OM-Key cylinder of particle 1. We propose that this gives an interpretation of the OM-Key cylinder mechanism as a Key sharing protocol \cite{menezes2018handbook}. If we consider the EPR version where we have two final observers forming the consensus view of the process then this consensus version is formed by using the OM-Key of the other final observer for "decrypting" their "private" or local OM-loops information. The end process of the OM-Key cylinder mechanism between both final observers will be the same state with opposite orientation. This is completely analog to the Private-Public Key protocols used in Informatics. We propose the OM-Key cylinder mechanism in entangled systems is indeed analog to a Private-Public Key protocol. More specifically a protocol using elliptic curves or higher order curves for decryption.  \\ \\

We know that if there was no entanglement box the OM-state to which we would collapse would be precisely $\lfloor (l_1 |_n)/2 \rfloor$. However, as in the EPR setup the OM-key needs first to "open" the curvature present in the entanglement box then now the OM-state to which each of the OM-mixed states will collapse will be $\lfloor ((l_1 - b) |_n)/2 \rfloor$. Where b is the scale of the OM-entanglement box. In the typical EPR setup n=2 and then this state to which we can collapse can be a OM-spin $ \frac{1}{2} $ state or a OM-spin $ - \frac{1}{2} $ which due to the uncertainty in the geometric side gives almost a $ 50\% $ probability for each value. Thus, the result in the OM-QM side is totally deterministic while at the same time mimicking an apparently totally random result (but just if we have uncertainty in the geometric side). This reproduces the identical absolute value of both measured OM-spins. Now let's see why we get for one particle OM-spin $ \frac{1}{2} $ state and OM-spin $ - \frac{1}{2} $ for the other particle. This can be seen easily in Fig......   The OM-Keys do rotate in opposite directions respect to each other due to the crossing (or crossings) present in the OM-entanglement box. Thus, we can consider the orientation of this rotation to give us the sign of the OM-base states to which we collapse. As both keys will "open" their respective OM-cylinders and half OM-entanglement boxes rotating exactly the same number of times in opposite directions, then the end orientation of both OM-keys will be exactly opposite. This concludes the mapping of the EPR experiment result to the OM-QM side. We find the result would be totally deterministic if we would not have uncertainty in the experimental space-temporal setup. \\ \\ 

Finally, we point out that similar as we did for the case of the OM-mixed state function Eq.\ref{55} we can get a general formula for the OM-entanglement curvature. For that let's recognize that the global topology of this two sided OM-key cylinder mechanism is that of a genus-2 torus. More precisely it is a torus with two holes and a connecting central region of length b (the entanglement box scale). We can see this visualization in the following figure

\begin{figure}[h]
	\centering
	\includegraphics[width=0.7\linewidth]{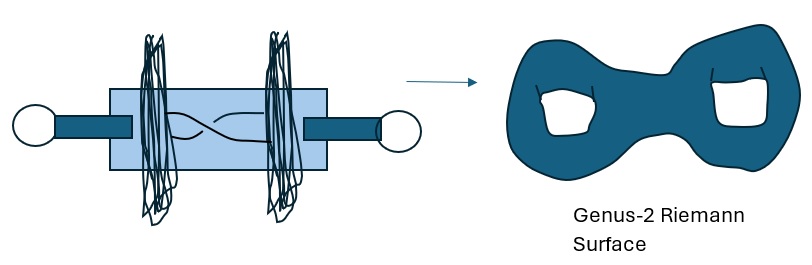}
	\caption[Figure 12]{The OM-RH space of two particle's as a genus-2 Riemann Surface}
\end{figure} 

If we extrapolate the conclusions we got for the case of one single OM-key cylinder we can conjecture that the entangled OM-state for two particles will be described by the Weirstrass function for a genus-2 Riemann Surface ($ \Phi_{genus-2} (z)$) \cite{england2012generalised} in this way 

\begin{equation}\label{70}
e^{vol_{e_1~e_2}(u)} = \frac{\Phi_{genus~2} (z)}{e^{vol_{s_1}(u)} ~ e^{vol_{s_2}(u)} ~e^{vol_{q_2}(t)} ~ e^{vol_{q_2}(t)}}
\end{equation}

where 

\begin{equation}\label{71}
\begin{array}{cl}
e^{vol_{e_1~e_2}(u)} & \mbox{is the OM-state entanglement part} \\
e^{vol_{s_1}(u)} & \mbox{is the OM-state part from the OM-petal for particle 1} \\
e^{vol_{s_2}(u)} & \mbox{is the OM-state part from the OM-petal for particle 2} \\
e^{vol_{q_1}(u)} & \mbox{is the OM-state part from the contractible OM-loops for particle 1} \\
e^{vol_{q_2}(u)} & \mbox{is the OM-state part from the contractible OM-loops for particle 2} 
\end{array}
\end{equation}

All the elements in the r.h.s. of this equation are known. Thus this can serve as a definition of OM-entanglement in OM-RH space. \\ \\

In a more general setup where we would have k geometrical constraints then we conjecture the OM-mixed state in OM-RH space will be  $ \Phi_{genus~k} $, the generalization of the Weirstrass function to a Rimemann surface of genus k. In [] we see how to construct such generalizations. These functions are in general modular forms over high dimensional spaces. The dimensionality of those spaces increases with the number of geometrical constraints in the QM side. Exploring how this increases in complexity and dimensionality impacts the OM-entanglement as in Eq.\ref{70} is an interesting future direction of this investigation. Also exploring in detail the OM-Key cylinder mechanism (that is the OM correspondent to the quantum state collapse mechanism) as a Public-Private Key protocol over elliptic curves or higher order curves is a potentially high value area for further development of the matters discussed in this section.  \\ \\

\section{Discussion and outlook}
OM-QM is the residual (reduced) Theory of QM when we modularize either the space-time background or the Hilbert space according to regions with observation or observed stated respectively. It is a mathematical framework living on a moduli space constructed via modularizaton. In this work we have extended OM-QM to include not only observation of space-time but also observation of quantum states. We have done that via the addition of a fourth step to the three ones that defined OM-space, the moduli space constructed via modularization of regions of space-time where observation was implied. This Step 4 states that after observation of a quantum state we should identify all the quantum states which are equivalent to the observed OM-base state. That equivalence relation is a way to pass from one OM-base state to another. We called that relation the OM-Key cylinder mechanism. This defines a product moduli space of OM-space and the modulirized OM-Hilbert space that we called  OM-Riemann-Hilbert space (OM-RH space). This product space is a the mathematical space on which the full OM-QM lives. It has the mathematical structure of a Riemann Surface. This allows us to make use of the very rich machinery of Riemann Surfaces for exploring OM-RH space and OM-QM defined on it. We found the OM-equivalents of quantum base states, mixed states and the Born rule. Using the visualization of the OM-Knot center region via  the OM-Key cylinder mechanism we found the OM-QM analog of the quantum state collapse process. In OM-QM there is a perfectly deterministic (but chaotic in the the OM-H sector of OM-RH) mechanism describing the state reduction process. We found an analytic way to induce this collapse process via a transformation in OM-RH space ($ u \longrightarrow \zeta(t) $) where u and t are complex and real componentes of a complex variable $ z=u + i~t $. The connections of Riemann Surfaces to several other mathematical objects (like Elliptic Curves and Elliptic Curve Encryption) give us interconnections among OM-QM objects which are not coming from the quantum side. One example of this is that we get an interpretation of the OM-Key cylinder mechanism as decryption in an Elliptic Curve  (ECC) framework. We found the OM-analog of entanglement between two particles in an EPR experiment setup. Its visualization is given by a pair of symmetric, interconnected OM-Key cylinder mechanisms. The details of the curvature joining of the OM-Keys and the OM-cylinders in this EPR setup shows that in a measurement in the EPR experiment the OM-keys are exchanged. They open the opposite cylinder instead of the one of their own particle. We know OM-QM is a purely mathematical framework. However we have conjectured that it can shed light on some aspects of the quantum side. OM-QM is constructed from QM but we can also get some feedback from OM-QM to QM. As an example of this we can mention that according to the ECC key exchange decryption protocol view of the EPR measurement there is no consensus view of the outcome of the EPR measurement until both particles are measured either by one single observer or via two different observers. The public key exchange between both observers is what allows the formation of a consensus view of the results of the experiment. That consensus view in EPR is that both observers will get the same end base state (with opposite sign). That is, the world view both observers observe is the same and unique and is there result of the ECC decryption protocol. This is completely analog o the use of ECC decryption in Informatics. In ECC encryption there are two observers with certain private internal states and certain public keys they can share with the other observer. Messages exchanges with meaning for one of the observers in its own internal view will look like entangled of encrypted to the other observer. In order for the other observer to get the same meaning of the view of the world both observers exchange their public keys and use them for encrypting and decrypting the observations. The modular mechanisms implicit in the ECC protocols cause that the end result meaning of the internal world views of both observers will be exactly the same. It has to be clear that the internal mechanisms of both observers can be completely different but the ECC protocol causes that both will understand encrypted information in the same way. In OM-QM the public keys are the OM-Keys, that is, the OM-petals. That means, the public keys are related to the geometric or space-time contents of the Theory.  By the other hand we saw that the exchange of public keys arises only because of the presence of the entanglement OM-skew regions of the OM-Knot. The possibility to have this OM-Key exchange is then tied to the presence of entanglement. We conjecture that the public key exchange in the Physical side will be related to very concrete space-time geometrical objects. To find which are those objects we can look at the Maldacena-Susskind ER=EPR framework \cite{susskind2016copenhagen}. In it, entanglement among identifiable regions of space time is analog to a General Relativity's Wormhole space-time geometry. We propose that as the exchange of public OM-Keys occurs via the region of entanglement among both EPR OM-Knots the analog of a GR Wormhole in OM-QM is this pair of OM-Keys inserted and locked in the OM-cylinder of the other particle. The plus we have in OM-QM is that we have an unified mathematical Theory for understanding (and visualizing) the dynamics of the effects of those OM-Keys over OM-quantum states. OM-RH space is the space where the OM-QM analog of ER=EPR is in reality operating in an unified way. Not with GR and QM as separated entities but instead as a single background space having the structure of a Riemann Surface. These relations of ER=EPR with OM-QM can be very interesting venues for future investigation\\ \\
Using the fact that OM-RH has the structure of a Riemann Surface we have proposed the Weirstrass Phi function is living in it is the general form of a OM-mixed state. Then the known differential equation for this function on the complexified space ($ z=u+i~t $) gives the dymanics of this OM-mixed state on diverse kind of processes which are in general not accessible to the realm of standard QM. We proved that the quantum state collapse is one of them. \\
We proposed the OM-states for two entangled particles can be described as the higher order Weirstrass for a genus-2 Riemann Surface. In general a quantum experimental setup containing N geometrical constraints and k entanglement regions will be represented in OM-QM by the higher order Weirstrass function over a Riemann Surface of genus-N and k connecting sections. One possible such higher genus Riemann Surface is a multi-Torus with N handles and k connecting sections among them. Investigating how this allows applying OM-QM to very diverse experiments and experimental setups is another potential future way for extend the scope of this work. As one possible example of this we will mention the following. In this work we analyzed the classical EPR setup which is totally symmetric respect to both involved particles. In a future work we intend to analyze also an asymmetric version of EPR experiment where one particle departs from the entanglement box without any further interaction until the moment it is measured. The other one however can be allowed to have several entanglements and geometrical constraints which, if they leave no final measurable trace (for example in scrambled or ultra-fast measurements) should have no effect in the final results of the classical EPR measurement in both particles. In this case OM-RH space for this system will be a Riemann Surface with one handle for the un-perturbed particle connected to a complex network of handles and connection sections for the other particle. The number of possible contractible OM-loops in both particles is also very different. How can we predict which will be the concrete OM-base state which will be measured ? In the symmetric EPR case we could predict exactly which is that measured OM-base state. In this very asymmetric case the answer is not so evident.  \\ \\ 
Finally, we think that there is much more to be learned from the higher order Weirstrass function as the OM-mixed state corresponding to general experimental setups. One can get explicit functions describing entanglement regions of the corresponding Riemann Surface and that could be mapped to explicit OM versions of GR Wormholes (OM-key regions of interaction with the OM-key cylinder). There are also many highly non-trivial relations between genus number, number of handle connections and the order of higher order Curves (of which Elliptic Curves are the lowest order example) which could give some insight on the structure of OM-QM and then also of QM and its relations to other Physical Theories as General Relativity. \\ \\
Finally we want to signal that our result that we can predict in a deterministic way exactly which concrete quantum base state we could get after a measurement (if we could control the geometry and number of quantum base states) is indeed an experimental prediction of OM-QM. In fact this is a prediction that the total random behavior of the outcome of a punctual quantum measurement is only a low dimensional illusion. In reality the result is totally fixed in OM-RH space. There could be experimental limits of setups in which this prediction could be falsified. We intend to explore this in future works.
 
 \bibliographystyle{unsrt}
 \bibliography{bibliogra1}

\begin{thebibliography}{10}

\bibitem{frugone2023quantum}
Jose A~Pereira Frugone.
\newblock Quantum mechanics on a background modulo observation.
\newblock {\em arXiv preprint arXiv:2311.12493}, 2023.

\bibitem{teleman2003riemann}
C~Teleman.
\newblock Riemann surfaces.
\newblock {\em Online Lecture Notes: http://math. berkeley.
  edu/teleman/math/Riemann. pdf}, 2003.

\bibitem{gajbhiye2011survey}
Samta Gajbhiye, Monisha Sharma, and Samir Dashputre.
\newblock A survey report on elliptic curve cryptography.
\newblock {\em International Journal of Electrical and Computer Engineering},
  1(2):195, 2011.

\bibitem{menezes2018handbook}
Alfred~J Menezes, Paul~C Van~Oorschot, and Scott~A Vanstone.
\newblock {\em Handbook of applied cryptography}.
\newblock CRC press, 2018.

\bibitem{england2012generalised}
Matthew England and Chris Athorne.
\newblock Generalised elliptic functions.
\newblock {\em Central European Journal of Mathematics}, 10:1655--1672, 2012.

\bibitem{rossler1976equation}
Otto~E R{\"o}ssler.
\newblock An equation for continuous chaos.
\newblock {\em Physics Letters A}, 57(5):397--398, 1976.

\bibitem{letellier2006inequivalent}
Christophe Letellier, Elise Roulin, and Otto~E R{\"o}ssler.
\newblock Inequivalent topologies of chaos in simple equations.
\newblock {\em Chaos, Solitons \& Fractals}, 28(2):337--360, 2006.

\bibitem{mazur2012primes}
Barry Mazur.
\newblock Primes, knots and po.
\newblock {\em Lecture notes}, 2012.

\bibitem{morishita2009analogies}
Masanori Morishita.
\newblock Analogies between knots and primes, 3-manifolds and number rings.
\newblock {\em arXiv preprint arXiv:0904.3399}, 2009.

\bibitem{morishita2011knots}
Masanori Morishita.
\newblock {\em Knots and primes: an introduction to arithmetic topology}.
\newblock Springer Science \& Business Media, 2011.

\bibitem{callahan1998hyperbolic}
Patrick~J Callahan and Alan~W Reid.
\newblock Hyperbolic structures on knot complements.
\newblock {\em Chaos, Solitons \& Fractals}, 9(4-5):705--738, 1998.

\bibitem{susskind2016copenhagen}
Leonard Susskind.
\newblock Copenhagen vs. everett, and er= epr.
\newblock {\em arXiv preprint arXiv:1604.02589}, 2016.

\end{thebibliography}

\end{document}